\documentclass[a4paper,11pt]{article}
\usepackage{pos}
\usepackage{float}
\usepackage{microtype}
\graphicspath{{figures/}}

\newcommand{\pt}{p_{\rm T}}
\newcommand{\GeVc}{\mathrm{GeV}/c}
\newcommand{\Dzero}{\mathrm{D}^{0}}
\newcommand{\Dstar}{\mathrm{D}^{*+}}
\newcommand{\Lambdac}{\Lambda_{\rm c}^{+}}
\newcommand{\Jpsi}{\mathrm{J}/\psi}
\newcommand{\psitwo}{\psi\mathrm{(2S)}}

\DeclareCaptionFont{compact}{\fontsize{9.5}{11}\selectfont}
\title{Charm and beauty production in pp collisions with ALICE}

\manuallySeparateAuthors
\author*[a]{Jianhui Zhu}
\author{ for the ALICE Collaboration}

\affiliation[a]{Institute of Modern Physics, Fudan University, 220 Handan Road, Shanghai, 200433, China}

\emailAdd{jianhui.zhu@cern.ch}

\abstract{
Measurements of charm and beauty production in proton--proton (pp) collisions provide crucial tests of perturbative quantum chromodynamics (pQCD) and unique access to heavy-quark hadronisation.
In this contribution, a selection of recent ALICE results, with emphasis on LHC Run~3 measurements, is presented. 
The transverse-momentum-differential production cross sections of prompt D mesons at $\sqrt{s}=5.36$ and 13.6~TeV and of fully reconstructed B mesons at $\sqrt{s}=13.6$~TeV are discussed together with comparisons to theoretical calculations.
The rapidity dependence of charm and beauty production and the $\rm D^{*+}/D^0$ and $\rm \Lambda_c^+/D^0$ yield ratios provide complementary constraints on heavy-quark production and hadronisation. Measurements of the fractions of ${\rm J}/\psi$ and $\psi({\rm 2S})$ originating from beauty-hadron decays, and of the inclusive $\psi({\rm 2S})$-to-${\rm J}/\psi$ production ratio, further constrain the interplay between perturbative heavy-quark production and non-perturbative bound-state formation.}

\FullConference{13th International Conference on Hard and Electromagnetic Probes of High-Energy Nuclear Collisions (Hard Probes 2026)\\
21--26 June 2026\\
Vanderbilt University, Nashville, Tennessee, USA}

\begin{document}
\maketitle

\linespread{1.07}\selectfont

\section{Introduction}
Measurements of heavy-flavour hadron production test quantum chromodynamics (QCD), with the heavy-quark masses providing hard scales for perturbative QCD (pQCD) calculations. In the factorisation approach, hadron cross sections are expressed as convolutions of parton distribution functions, partonic cross sections and non-perturbative fragmentation functions~\cite{Collins}. These fragmentation functions are usually constrained using $\mathrm{e^+e^-}$ data, under the assumption that they are universal between collision systems. Hadron yield ratios probe hadronisation with reduced sensitivity to initial-state and partonic-production effects.

Charm and beauty hadrons and their charge conjugates are reconstructed from their decay products in the ALICE detector. In the following, ``prompt'' charm hadrons include feed-down from promptly produced charm states, while ``non-prompt'' charm hadrons originate from beauty-hadron decays. The upgraded ALICE detector and Run 3 data samples extend these studies to low transverse momentum ($\pt$) and allow the full reconstruction of beauty-hadron decays. 

\section{Results}
\subsection{D- and B-meson production}
\begin{figure}[H]
\centering
\includegraphics[height=50mm]{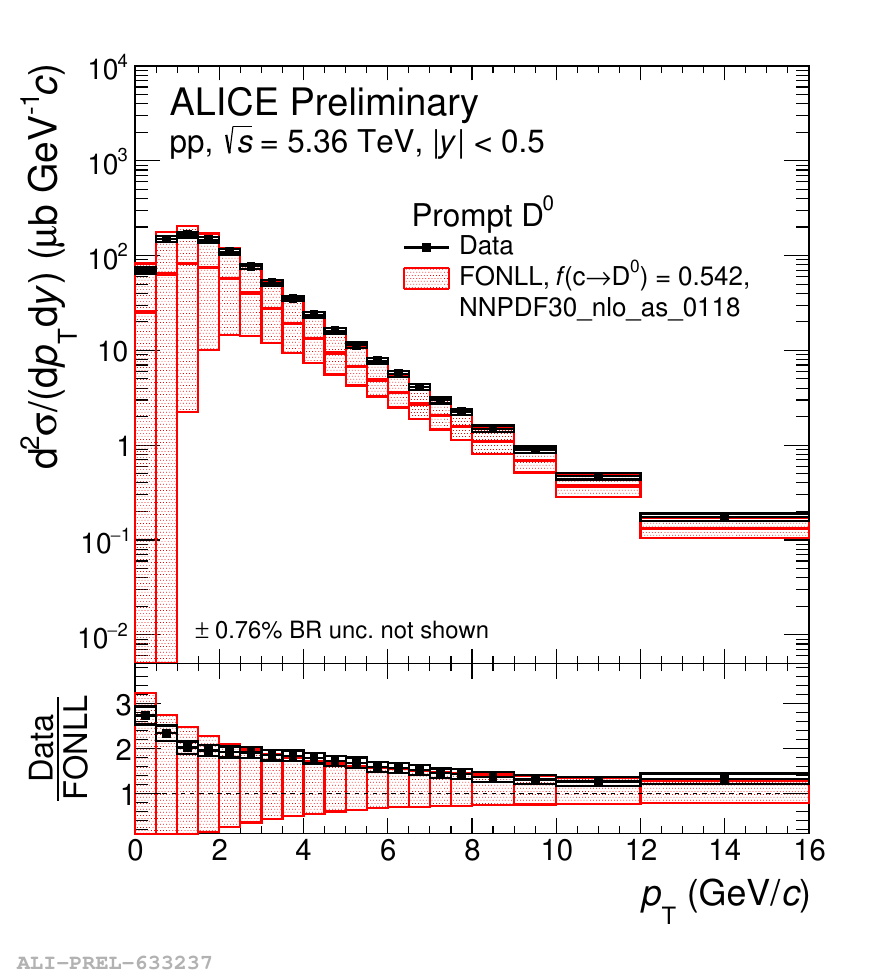}\hspace{4mm}%
\includegraphics[height=49mm]{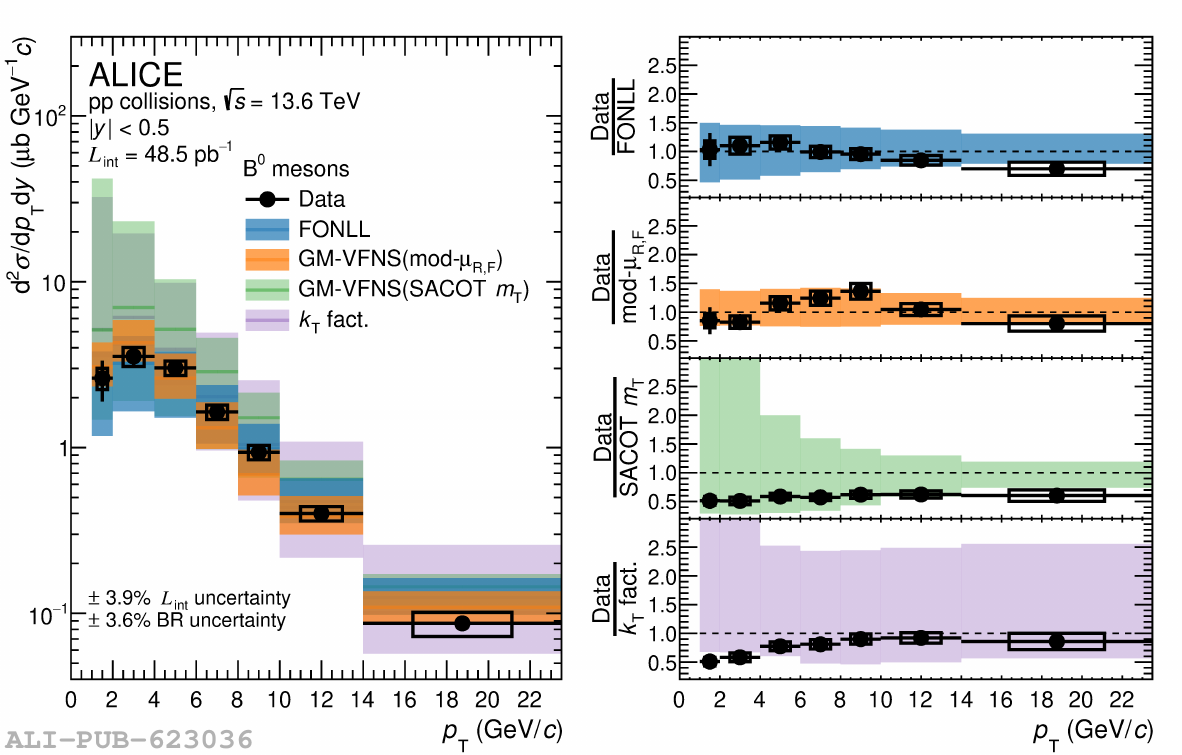}
\caption{Left: preliminary prompt $\Dzero$ cross section at $\sqrt{s}=5.36$~TeV, compared with FONLL~\cite{FONLL}. Right: $\mathrm{B^0}$ cross section at $\sqrt{s}=13.6$~TeV, compared with pQCD calculations~\cite{Bzero}. Both measurements are performed in pp collisions at $|y|<0.5$.}
\label{fig:cross}
\end{figure}

The preliminary prompt $\Dzero$ cross sections at $\sqrt{s}=5.36$ and 13.6~TeV and the $\Dstar$ cross section at 13.6~TeV are measured at $|y|<0.5$. The data are compatible with fixed-order plus next-to-leading-logarithm (FONLL) pQCD calculations~\cite{FONLL} and lie near the upper edge of their uncertainty bands, as shown for $\Dzero$ at 5.36~TeV in Fig.~\ref{fig:cross} (left). The experimental uncertainties are substantially smaller than the theoretical ones.

The first ALICE measurement of fully reconstructed $\mathrm{B^0}$ mesons, via $\mathrm{B^0}\to\mathrm{D^-}\pi^+$, extends down to $\pt=1~\GeVc$ at $|y|<0.5$ in pp collisions at $\sqrt{s}=13.6$~TeV~\cite{Bzero}. FONLL~\cite{FONLL}, general-mass variable-flavour-number-scheme (GM-VFNS)~\cite{GMVFNS}, and $k_{\rm T}$-factorisation~\cite{k_T} calculations describe the cross section within uncertainties (Fig.~\ref{fig:cross}, right). The $\mathrm{B^0}$ data also constrain phenomenological models~\cite{Bzero}. TAMU~\cite{TAMU} describes the measurement within uncertainties, while Catania underestimates it at $\pt<6~\GeVc$ and EPOS4HQ~\cite{EPOS4HQ} overestimates it at $\pt>10~\GeVc$. AMPT~\cite{AMPT} describes the data for a beauty-quark mass parameter of $6.6~\mathrm{GeV}/c^2$, but overestimates them for $4.8~\mathrm{GeV}/c^2$. The preliminary $\mathrm{B^{\pm}}$ measurement via $\mathrm{B^{\pm}}\to\Jpsi(\to\mathrm{e^+e^-})\mathrm{K^{\pm}}$ down to $\pt=2~\GeVc$ further supplements the beauty measurements at midrapidity.

Mid-to-forward-rapidity cross-section ratios are obtained from the ALICE $\mathrm{B^0}$ measurement at 13.6~TeV and the LHCb $\mathrm{B^+}$ measurements at 13~TeV~\cite{Bzero}. The ratios increase with $\pt$, with a steeper increase at larger forward rapidities, indicating harder spectra at midrapidity. The $\Dzero$ ratios show a similar trend. FONLL~\cite{FONLL} and next-to-next-to-leading-order plus next-to-next-to-leading-logarithm (NNLO+NNLL) predictions~\cite{NNLO} qualitatively describe this behaviour, with the latter predicting larger ratios. Both calculations account for the different collision energies.

\subsection{Charm meson-to-meson and baryon-to-meson yield ratios}

\begin{figure}[H]
\centering
\includegraphics[height=50mm]{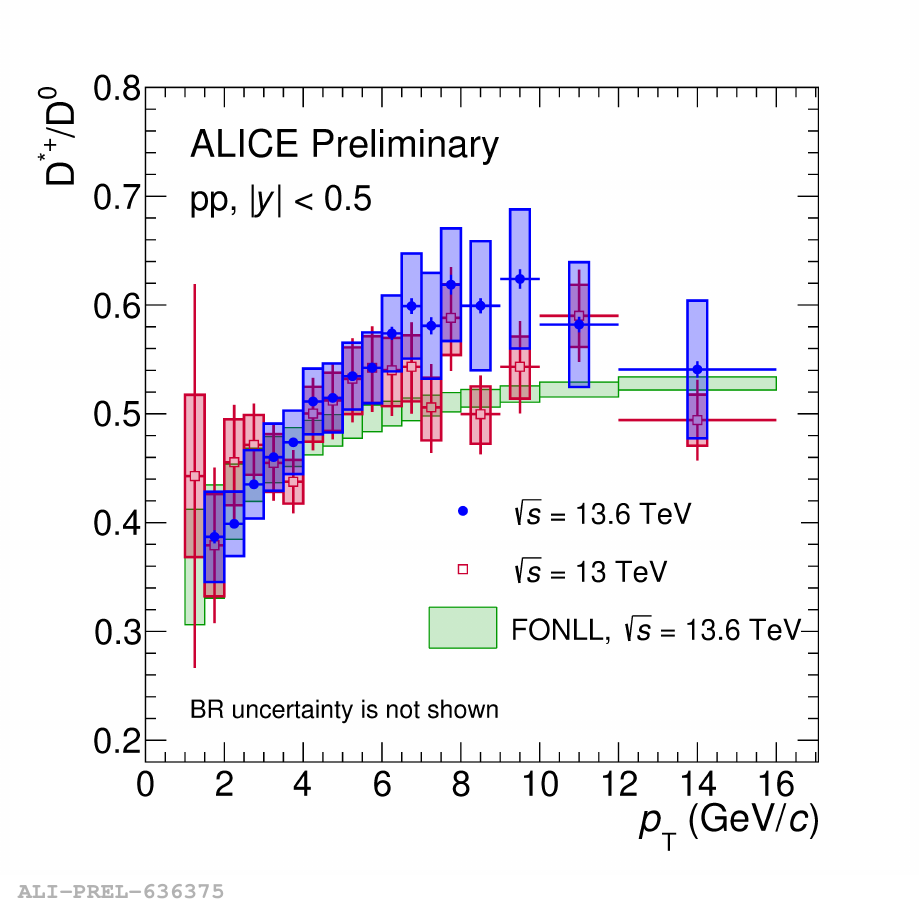}\hspace{5mm}%
\includegraphics[height=47mm]{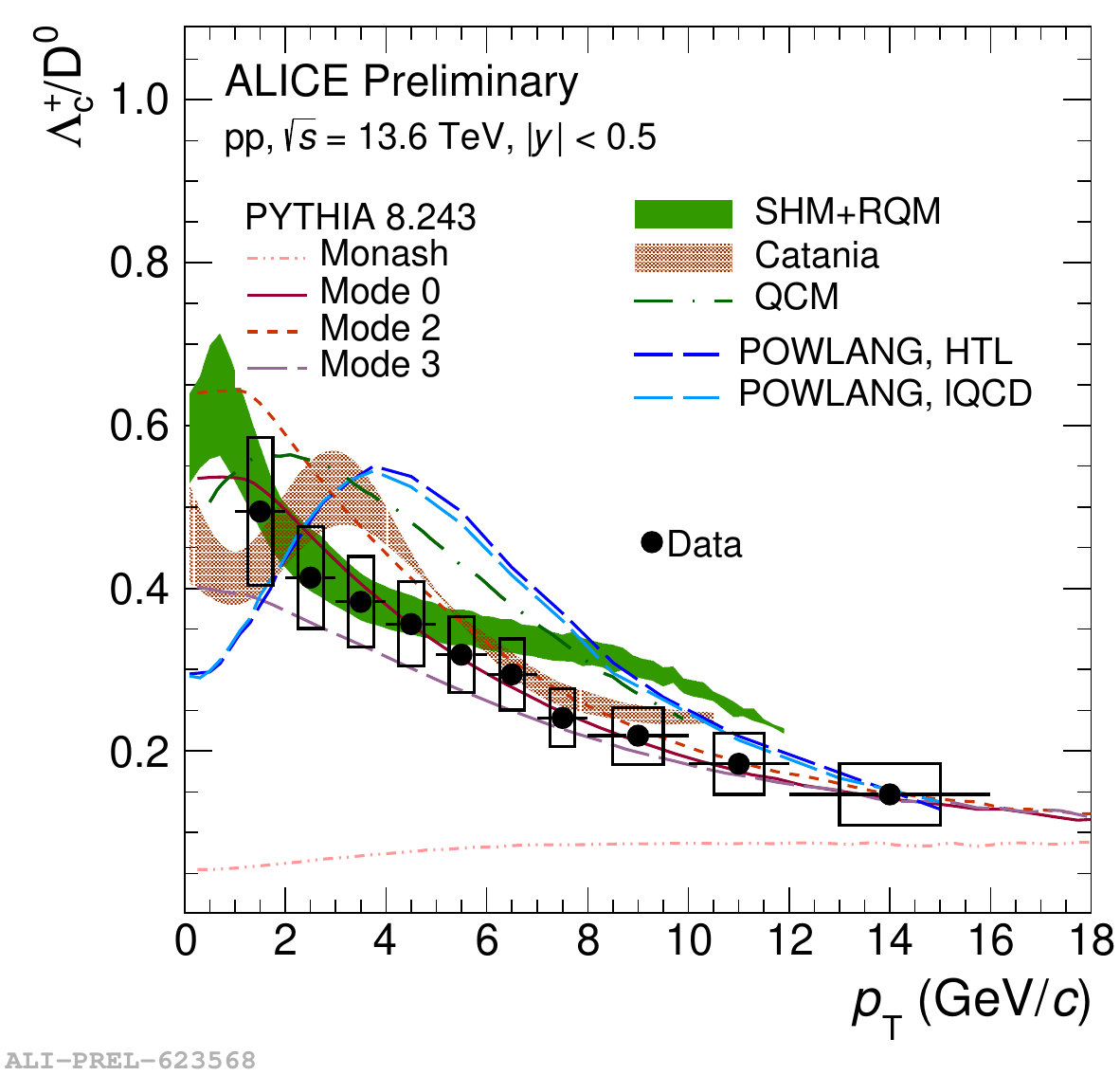}
\caption{Preliminary charm-hadron yield ratios in pp collisions at $\sqrt{s}=13.6$~TeV and $|y|<0.5$. Left: $\Dstar/\Dzero$, compared with the Run~2 result at 13 TeV and with FONLL~\cite{FONLL}. Right: $\Lambdac/\Dzero$, compared with hadronisation models.}
\label{fig:ratios}
\end{figure}

The preliminary $\Dstar/\Dzero$ yield ratio at $\sqrt{s}=13.6$~TeV  (Fig.~\ref{fig:ratios}, left) is compared with the Run~2 result at 13~TeV. The Run~3 measurement is compatible with the Run~2 result within uncertainties and has better statistical precision. The ratio increases with $\pt$ up to about $8~\GeVc$ and is reproduced by FONLL. The preliminary $\Lambdac/\Dzero$ yield ratio at $\sqrt{s}=13.6$~TeV is consistent with the Run~2 result at 13~TeV (Fig.~\ref{fig:ratios}, right). The enhancement at low and intermediate $\pt$ is underestimated by PYTHIA 8 with the Monash tune, which uses fragmentation constrained by $\mathrm{e^+e^-}$ data. Models incorporating colour reconnection beyond the leading-colour approximation~\cite{CR}, coalescence (Catania, QCM and POWLANG)~\cite{Catania,QCM,POWLANG}, or feed-down from additional excited charm baryons in statistical hadronisation with a relativistic-quark-model spectrum (SHM+RQM)~\cite{SHM} capture aspects of the enhancement. The data do not uniquely select a hadronisation mechanism.

\subsection{Charmonium production and non-prompt fractions}
\begin{figure}[H]
\centering
\includegraphics[height=50mm]{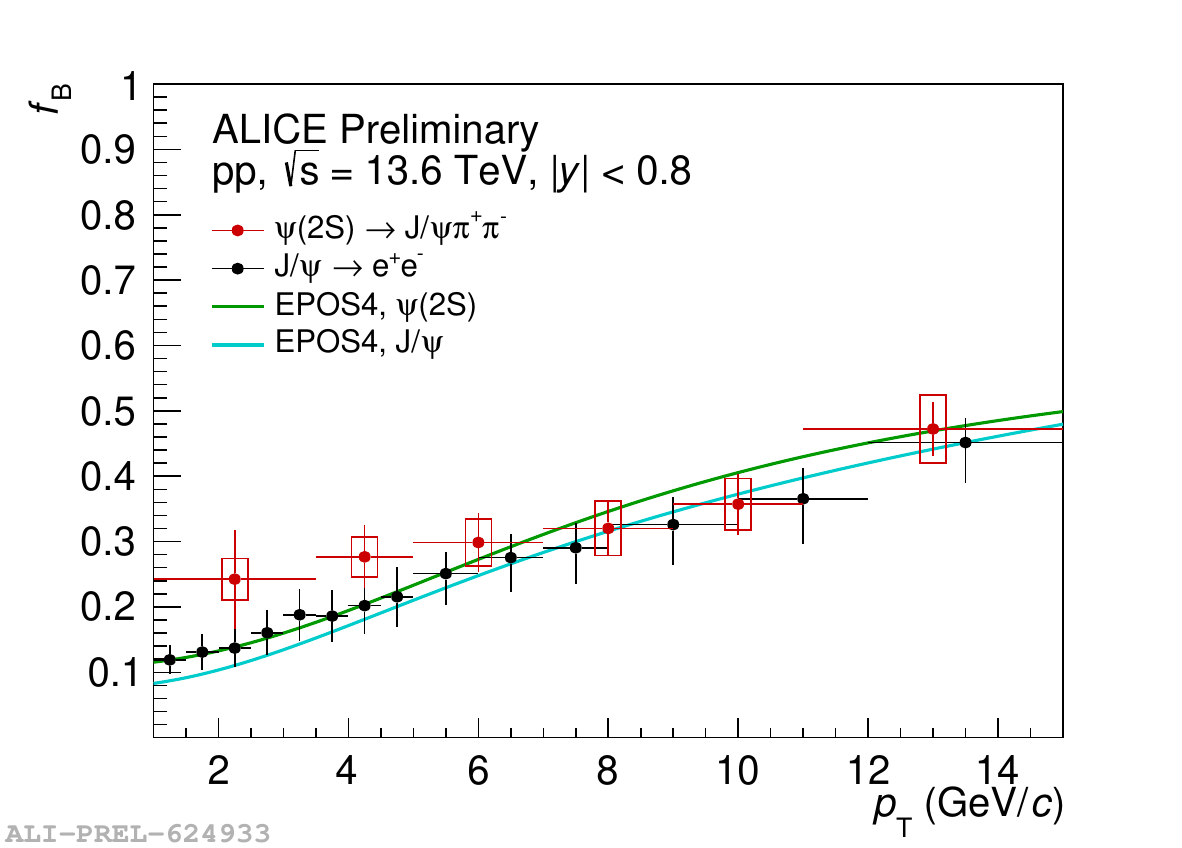}\hspace{5mm}%
\includegraphics[height=50mm]{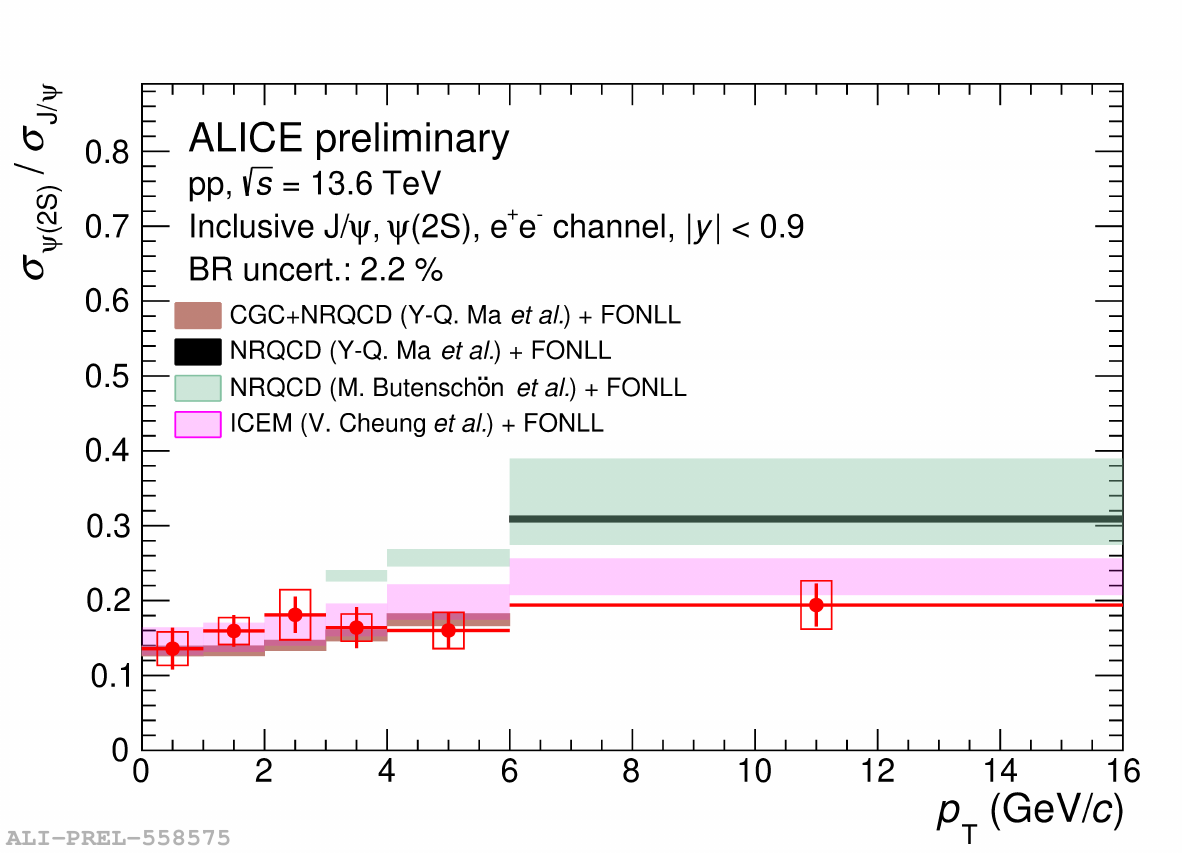}
\caption{Preliminary charmonium measurements in pp collisions at $\sqrt{s}=13.6$~TeV. Left: non-prompt fractions of $\mathrm{J}/\psi$ and $\psi(2\mathrm{S})$ at $|y|<0.8$, compared with EPOS4. Right: inclusive $\psi(2\mathrm{S})$-to-$\mathrm{J}/\psi$ cross-section ratio at $|y|<0.9$, compared with NRQCD, CGC+NRQCD and ICEM predictions, including non-prompt contributions from FONLL~\cite{FONLL}.}
\label{fig:charmonia}
\end{figure}

Charmonium production probes heavy-quark-pair production and bound-state formation. The non-prompt fraction, $f_{\rm B}=\sigma_{\rm non\text{-}prompt}/\sigma_{\rm inclusive}$, quantifies the contribution from beauty-hadron decays. For $\Jpsi$ at $\sqrt{s}=13.6$~TeV, it increases with $\pt$, a trend described by PYTHIA~8 with colour-reconnection Mode~2. It also increases with event activity, particularly at low $\pt$. The first low-$\pt$ measurement of the non-prompt $\psitwo$ fraction at midrapidity uses $\psitwo\to\Jpsi\pi^+\pi^-$, with $\Jpsi\to\mathrm{e^+e^-}$. The non-prompt fraction of $\psitwo$ shows a hint of being larger than that of $\Jpsi$ at low $\pt$ (Fig.~\ref{fig:charmonia}, left), qualitatively consistent with the larger prompt feed-down contribution to $\Jpsi$ from excited charmonia. The EPOS4~\cite{EPOS4HQ} predictions shown reasonably describe the $\pt$ dependence of the non-prompt fractions for both states.

The inclusive $\psitwo$-to-$\Jpsi$ cross-section ratio shows a mild $\pt$ dependence and no significant dependence on collision energy or rapidity within uncertainties. The Run~3 measurement has better statistical precision than the Run~2 result. The improved colour evaporation model (ICEM)~\cite{ICEM_1, ICEM_2} describes the ratio at midrapidity and forward rapidity, whereas the nonrelativistic-QCD (NRQCD) predictions considered~\cite{NRQCD_1, NRQCD_2} overestimate it at high $\pt$ at midrapidity. The comparisons in Fig.~\ref{fig:charmonia} (right) include non-prompt contributions from FONLL~\cite{FONLL}.

\section{Summary}
The ALICE Run 3 measurements test pQCD through charm and beauty production at low $\pt$. The enhanced $\Lambdac/\Dzero$ ratio challenges fragmentation models tuned to $\mathrm{e^+e^-}$ collision data. Charmonium production ratios and non-prompt fractions further test charmonium formation within the NRQCD framework and constrain beauty production, respectively. Measurements in additional fully reconstructed beauty-hadron decay channels will further tighten these constraints.

\medskip
\noindent\textbf{Acknowledgments.}
This work was supported by the National Natural Science Foundation of China (NSFC) under Grant No.~12675171 and by the Ministry of Science and Technology of China (MOST) under Grant Nos.~2024YFA1610801 and 2024YFA1610702.



\begin{thebibliography}{99}
\bibitem{Collins} J.~C.~Collins, D.~E.~Soper and G.~Sterman, \textit{Adv.Ser.Direct. High Energy Phys.} \textbf{5} (1989) 1-91.
\bibitem{FONLL} M.~Cacciari et al., \textit{JHEP} \textbf{10} (2012) 137.
\bibitem{Bzero} ALICE Collaboration, {arXiv:2603.18904} [hep-ex].
\bibitem{GMVFNS} M.~Benzke et al., \textit{Eur.Phys.J.C} \textbf{79} (2019) 814.
\bibitem{k_T} F. Barattini et al., \textit{JHEP} \textbf{05} (2025) 115.
\bibitem{TAMU} M.~He and R.~Rapp, \textit{Phys. Rev. Lett.} \textbf{131} (2023) 012301.
\bibitem{EPOS4HQ} J.~Zhao et al., \textit{Phys. Rev. C} \textbf{110} (2024) 024909.
\bibitem{AMPT} J.~He et al., \textit{Phys. Rev. D} \textbf{113} (2026) 034008.
\bibitem{NNLO} M.~Czakon et al., \textit{Phys. Rev. Lett.} \textbf{135} (2025) 161903.
\bibitem{CR} J.~R.~Christiansen and P.~Z.~Skands, \textit{JHEP} \textbf{08} (2015) 003.
\bibitem{Catania} V.~Minissale, S.~Plumari and V.~Greco, \textit{Phys. Lett. B} \textbf{821} (2021) 136622.
\bibitem{QCM} J.~Song et al., \textit{Eur. Phys. J. C} \textbf{78} (2018) 344.
\bibitem{POWLANG} A.~Beraudo et al., \textit{Phys. Rev. D} \textbf{109} (2024) L011501.
\bibitem{SHM} M.~He and R.~Rapp, \textit{Phys. Lett. B} \textbf{795} (2019) 117-121.
\bibitem{ICEM_1} Y.~Q.~Ma and R.~Vogt, \textit{Phys. Rev. D} \textbf{94} (2016) 114029.
\bibitem{ICEM_2} V.~Cheung and R.~Vogt, \textit{Phys. Rev. D} \textbf{98} (2018) 114029.
\bibitem{NRQCD_2} M.~Butenschoen and B.~A.~Kniehl, \textit{Phys. Rev. Lett.} \textbf{106} (2011) 022003.
\bibitem{NRQCD_1} Y.~Q.~Ma, K.~Wang and K.~T.~Chao, \textit{Phys. Rev. Lett.} \textbf{106} (2011) 042002.
\end{thebibliography}

\begingroup
\fontsize{9.5}{8.4}\selectfont

\endgroup

\end{document}